\documentclass[aps,prb,twocolumn,superscriptaddress]{revtex4-1}

\usepackage{graphicx}
\usepackage{subfigure}
\usepackage{dcolumn}
\usepackage{bm}
\usepackage{amsmath}
\usepackage{amssymb}
\usepackage{latexsym}
\usepackage{epsfig}
\usepackage{amsbsy}
\usepackage{array}
\usepackage{amssymb}
\usepackage{setspace}

\begin{document}
\title{Modulation of pairing symmetry with bond disorder in unconventional superconductors}
\author{Yao-Tai Kang}
\affiliation{State Key Laboratory of Optoelectronic Materials and Technologies, School of Physics, Sun Yat-Sen University, Guangzhou 510275, China}
\author{Wei-Feng Tsai}
\email{weifengt@gmail.com}
\affiliation{Department of Physics, National Sun Yat-sen University, Kaohsiung 80424, Taiwan}
\affiliation{State Key Laboratory of Optoelectronic Materials and Technologies, School of Physics, Sun Yat-Sen University, Guangzhou 510275, China}
\author{Dao-Xin Yao}
\email{yaodaox@mail.sysu.edu.cn}
\affiliation{State Key Laboratory of Optoelectronic Materials and Technologies, School of Physics, Sun Yat-Sen University, Guangzhou 510275, China}

\newcommand{\br}{\mathbf{r}}
\newcommand{\brprime}{{\mathbf{r}^\prime}}
\newcommand{\bk}{\mathbf{k}}
\newcommand{\bkprime}{{\mathbf{k}^\prime}}
\newcommand{\bq}{\mathbf{q}}
\newcommand{\hx}{\hat{x}}
\newcommand{\hy}{\hat{y}}


\begin{abstract}
We study a two-orbital $t$-$J_1$-$J_2$ model, originally developed to describe iron-based superconductors at low energies, in the presence of bond disorder (via next-nearest-neighbor $J_2$-bond dilution). By using the Bogoliubov--de Gennes approach, we self-consistently calculate the local pairing amplitudes and the corresponding density of states, which demonstrate a change of dominant pairing symmetry from $s_\pm$ wave to $d$ wave when increasing disorder strength as long as $J_1\lesssim J_2$. Moreover, the combined pairing interaction and strong bond disorder lead to the formation of $s_\pm$ wave ``islands'' with length scale of the superconducting coherence length embedded in a $d$ wave ``sea.'' This picture is further complemented by the disorder-averaged pair-pair correlation functions, distinct from the case with potential disorder, where the ``sea'' is insulating. Due to this inevitable formation of spatial inhomogeneity, the superconducting $T_c$ determined by the superfluid density $\rho_s(T)$ obviously deviates from the value predicted by the conventional Abrikosov-Gorkov theory, where the pairing amplitudes are viewed as uniformly suppressed as the disorder increases.
\end{abstract}
\maketitle

\section{Introduction}
Studying disorder effects in superconductors (SCs) is usually beneficial, though not in a direct manner, for unveiling their underlying pairing mechanism. An economic early indicator of the unconventional nature of a superconductor, for instance, can be the sensitivity of the superconducting transition temperature ($T_c$) to an amount of finite disorder.\cite{Anderson59,Abrikosov60,Sigrist91,Balatsky06,Alloul09} In addition, most unconventional SCs, such as cuprates\cite{Keimer15} and iron-based SCs\cite{Stewart11,Dai12,Chen14}, become superconducting after doping, which naturally introduces certain types of disorder in the materials. These materials are often complicated in composition and are intermediately or strongly coupled systems, causing a rather complex phase diagram with intertwined orders in which disorder might play a role.\cite{Fradkin15} With substantial amounts of disorder, a SC may even undergo a zero-temperature quantum phase transition to another superconducting phase with distinct pairing symmetry\cite{Ng09,Kontani10,Efremov11,Spivak08,Spivak09,Vojta05} or to a non-superconducting one.\cite{Fisher90,Dubi07}

From a modeling point of view, a real disorder environment in a system may be simulated by adding either random one-body or random many-body interactions. Both types of interactions include the variations of either amplitude or phase and can be further classified by the interaction range, i.e., short range or long range. For instance, in Zn-doped cuprates like YBa$_2$(Cu$_{1-x}$Zn$_x$)$_3$O$_{6.9}$ such a disorder effect could be represented by a random set of scalar impurity potentials with a finite range.\cite{Franz1997} In fact, disorder effects resulting from random one-body potentials in SCs are widely discussed,\cite{Franz1997,XiangTao1995,Ghosal2001,CHENhua2013} while, in contrast, those from random many-body interactions are relatively less studied in a systematic way.\cite{Josephson,Halley90,Parcollet99,Otsuki13,Liang15}

The discovery of iron-based SCs has enriched the unconventional SC physics in several aspects. One particularly interesting aspect is that the isovalent doping in both 1111 and 122 materials, which is theoretically predicted not to change electron density significantly,\cite{Coldea2008,Kasahara2010,Rullier2010} can suppress the stripe antiferromagnetic (AFM) order in the parent compounds.\cite{cruz2010,Kasahara12} Moreover, it even induces superconductivity after intermediate doping in 122 materials;\cite{Jiang09,Nakai2010a} in some cases, there have been observed nodal structures in the superconducting gap,\cite{Fletcher2009,Hicks09,Hashimoto2010,Nakai2010b,J.S.Kim2010} distinct from the fully gapped one upon charge doping. These findings may provide a possible playground to study random many-body interactions in SCs through the following intuition, provided the strong-coupling picture is used.\cite{seo2008} For instance, a common consensus in  BaFe$_2$(As$_{1-x}$P$_x$)$_2$ is that both magnetism and superconductivity should occur in the FeAs plane, at which Fe atoms themselves form a square lattice with As atoms sitting above and below each plaquette center of the lattice alternately. Assuming that the stripe AFM order arises from the competition between nearest-neighbor (NN) $J_1$ and next-nearest-neighbor (NNN) $J_2$ exchange interactions on the square lattice, the (random) substitution of As by P could result in two leading effects to disorder the system: one is to mainly suppress NNN $J_2$ exchange interactions\cite{Kuroki09,Mizuguchi10} and the other is to introduce a scalar potential at each plaquette center.

Thus, motivated by the experiments done in isovalent-doping iron-based SCs, we study the effects of purely exchange-interaction disorder (random $J_2$-bond dilution; see Sec. II for the definition), from ``weak'' to ``strong'' (i.e., $x$ from 0 to 1), in a two-dimensional (2D), two-orbital $t$-$J_1$-$J_2$ model.\cite{seo2008,Si08} Although this model was originally developed to describe the low-energy physics of iron-based SCs, we simply consider a relatively ideal situation, to make our study be a topic of broad interest, but do not intend to claim it directly applicable to a certain real material.

When the bond disorder is ``strong'' with $0\ll x<1$ in a SC with short coherence length $\xi$, the standard theories for dirty SCs, valid as long as the electron mean free path $l\gg \xi> k_F^{-1}$, by Anderson\cite{Anderson59} and by Abrikosov and Gorkov\cite{Abrikosov60} (AG) are expected to be insufficient to account for the disorder effects. Therefore, we address these issues by using self-consistent Bogoliubov--de Gennes (BdG) formulation with the emphasis on the spatial inhomogeneity for the pairing amplitudes. Significantly, we find the following:
(1) As $J_1\lesssim J_2$, the pairing symmetry of our model at zero temperature ($T=0$) is modulated from $s_{x^2y^2}$-wave to $d_{x^2-y^2}$-wave symmetry when the ``disorder strength'' $x$ becomes greater than $x_c$; this phenomenon is further confirmed by showing the electron density of states as a function of $x$.
(2) The $T=0$ spectral gap $E_{gap}$ decreases following the same trend of the position/disorder averaged pairing amplitudes, $\overline{\Delta}_{s_{x^2y^2}}$ and $\overline{\Delta}_{d_{x^2-y^2}}$, as $x$ grows.
(3) When $x$ is large, the combined pairing interaction and the $J_2$-bond dilution disorder lead to the formation of $s_{x^2y^2}$-wave ``islands'' of length scale $O(\xi)$ embedded in a $d_{x^2-y^2}$-wave ``sea''; this picture is complemented by the disorder-averaged pair-pair correlation functions.
(4) Due to this inevitable formation of spatial inhomogeneity, the superconducting $T_c$ determined by the superfluid density $\rho_s(T)$ obviously deviates from the value predicted by the AG theory, where the pairing amplitudes are viewed as uniformly suppressed as the disorder increases.

The rest of the paper is organized as follows. In Sec. II, we describe our model Hamiltonian and briefly sketch the numerical method we used. We then demonstrate our numerical results in Sec. III to show the modulation of the pairing symmetry with bond disorder. Several disorder-dependent physical quantities are presented to assist the understanding of this type of disorder, such as local pairing amplitudes, density of states, spectral gap, and superfluid density. In Sec. IV, we repeat all the calculations in Sec. III by taking away all $J_1$ exchange interactions to sharpen the effects due to $J_2$-bond dilution. Finally, we conclude our findings in terms of a few remarks and a summary.

\section{Model and Self-consistent BdG theory}
Our adopted model Hamiltonian to capture the low-energy physics in {\it clean} iron-based superconductors is the so-called $t$-$J_1$-$J_2$ model developed in Ref. \onlinecite{seo2008}, which has been further justified by functional renormalization group study.\cite{Zhai09} Explicitly, $H=H_0+H_{int}$ and the noninteracting part reads
\begin{eqnarray}
	H_0 = \sum_{\br\br^\prime}\sum_{\alpha\beta}\sum_{\sigma}
        \left(t_{\br\br^\prime}^{\alpha\beta} c_{\br\alpha\sigma}^{\dagger}c_{\br^\prime\beta\sigma} + H.c.\right)
        - \mu\sum_{\br\alpha} n_{\br\alpha}, \nonumber  \\
	\label{eq:h0}
\end{eqnarray}
where $c_{\br\alpha\sigma}^\dag \left(c_{\br\alpha\sigma}\right)$ creates (annihilates) an electron of the $\alpha$-orbital with spin $\sigma$ ($\alpha=1,2$ for two degenerate $d_{xz}$ and $d_{yz}$ orbitals, respectively) at site $\br$. $n_{\br\alpha}$ is the local electron density operators of the $\alpha$-orbital.

The normal-state Fermi surfaces in the unfolded Brillouin zone (one iron per unit cell) can be reasonably produced by setting $t_{\br,\br+\hx}^{11}=t_{\br,\br+\hy}^{22}=-1.0$,
$t_{\br,\br+\hx}^{22}=t_{\br,\br+\hy}^{11}=1.3$,  $t_{\br,\br\pm\hx\pm\hy}^{11/22}=-0.85$, $t_{\br,\br+\hx-\hy}^{12/21}=-0.85$, $t_{\br,\br+\hx+\hy}^{12/21}=0.85$, and other hopping parameters as zero, where $\hx$ and $\hy$ are unit vectors along the axes. For simplicity, we will take $|t_{\br,\br+\hx}^{11}|=1$ as our energy units, lattice constant $a\equiv 1$, and set chemical potential $\mu=1.8$, corresponding to electron density $n_e\approx 2.18$.

The interacting part includes several terms as follows,
\begin{eqnarray}
    H_{int} &=& \sum_{\langle\br\brprime\rangle}\sum_{\alpha}
    J_{1}(\br,\brprime)(\mathbf{S}_{\br\alpha}\cdot
    \mathbf{S}_{\brprime\alpha}-n_{\br\alpha}
    n_{\brprime\alpha}) \nonumber \\
    &+& \sum_{\langle\langle\br\brprime\rangle\rangle}\sum_{\alpha}
    J_{2}(\br,\brprime)(\mathbf{S}_{\br\alpha}\cdot
    \mathbf{S}_{\brprime\alpha}-n_{\br\alpha}
    n_{\brprime\alpha}) \nonumber \\
    &+& \cdots,
\end{eqnarray}
where
$\mathbf{S}_{\br\alpha}=c^\dagger_{\br,\alpha,\sigma}\vec{\sigma}_{\sigma\sigma^\prime}c_{\br,\alpha,\sigma^\prime}$
and $n_{\br\alpha}$ are the local spin and density operators with orbital index $\alpha=1,2$. $\langle \br\brprime\rangle$ and $\langle\langle \br\brprime\rangle\rangle$ denote NN and NNN pairs of sites, respectively, and thus the first two
terms represent intraorbital exchange interactions. In addition, ``$\cdots$'' represents our ignored interorbital exchange and Hund's coupling terms, which are shown to be unimportant in determining the pairing symmetry of the SC state in this model.

On a square lattice with $N=N_x\times N_y$ sites, the exchange couplings $J_2(\br,\brprime)$ are taken to be zero when the (diagonal) bonds $(\br\brprime)$ cross the randomly selected $xN$ plaquettes. In this way we introduce the ``disorder'' into our otherwise clean system. This kind of bond disorder is a quantum analog of bond-dilute Ising models\cite{Zobin78,Mina93} while it is rarely considered in superconducting systems. Physically, these selected plaquettes might represent the situation where the central atoms As are isovalently replaced by the atoms P, causing strong suppression of $J_2$ bonds in plaquettes. For simplicity, we further assume that the
exchange couplings $J_1(\br,\brprime)=J_1$ are unaffected. A more delicate choice for $J_1(\br,\brprime)$ will be discussed later.

Following Ref.~\onlinecite{seo2008}, we assume that the superconductivity of the system can be reliably captured by mean-field approximation as long as the exchange interactions are small compared to the bandwidth ($\sim12|t_1|$). The most dominant SC order parameters in our study would generally have the following symmetry form factor: $\Delta_\alpha=a_\alpha \cos k_x\cos k_y-(-1)^\alpha b_\alpha(\cos k_x-\cos k_y)$. The relative sign between $a_1(b_1)$ and $a_2(b_2)$ determines which irreducible representation the pairing symmetry belongs to, namely, $A_{1g}$ for the plus sign and $B_{1g}$ for the minus sign, given $D_{4h}$ point group symmetry of our model. In the presence of disorder, we define the (local) real-space pairing amplitude for each orbital $\alpha$ as
\begin{equation}
    \Delta_{\alpha}(\br,\br+\delta)=-J_l(\br,\br+\delta)\langle c_{\br,\alpha,\downarrow}c_{\br+\delta,\alpha,\uparrow}
    \rangle
\end{equation}
with $\delta=\pm\hat{x},\pm\hat{y}$ for NN pairing ($l=1$) and $\delta=\pm\hat{x}\pm\hat{y}$ for NNN pairing ($l=2$). The mean-field Hamiltonian of $H$ is then written as
\begin{eqnarray}
    H^{MF} = H_0 + \sum_{\br,\delta,\alpha}\Delta^*_{\alpha}(\br,\br+\delta)c_{\br,\alpha,\downarrow}
    c_{\br+\delta,\alpha,\uparrow}+H.c.  \label{eq:mfH}
\end{eqnarray}

Within the BdG formalism, we diagonalize the {\it quadratic}, mean-field Hamiltonian (\ref{eq:mfH}) through the BdG equation,
\begin{eqnarray}
    \left(\begin{array}{cc}
        \hat{K}_{\br\alpha\brprime\beta} & \hat{\Delta}_{\alpha\beta}                  \\
        \hat{\Delta}^*_{\alpha\beta} &  -\hat{K}^*_{\br\alpha\brprime\beta}
        \end{array}
        \right)\left(\begin{array}{c}
        u_{\brprime\beta}^n    \\
        v_{\brprime\beta}^n
        \end{array}
        \right)=E_n\left(\begin{array}{c}
        u_{\br\alpha}^n \\
        v_{\br\alpha}^n
    \end{array}\right),
    \label{eq:bdg}
\end{eqnarray}
with $\hat{K}_{\br\alpha\brprime\beta} = -t_{\br\alpha\brprime\beta} -\mu\delta_{\br\brprime}\delta_{\alpha\beta}$.
The relation between Bogoliubov quasiparticle operators $\gamma$ and electron operators is
$c_{\br\alpha\sigma}=\sum_{n}(u^n_{\br\alpha}\gamma_{n\sigma}-\sigma v^{n*}_{\br\alpha}\gamma^\dagger_{n\bar{\sigma}})$,
and hence combining with the definition, for instance, of the $s$-wave $\cos k_x\cos k_y$ SC order parameter, this gives rise to the following self-consistent
conditions,
\begin{eqnarray}
    \Delta_{\alpha}(\br,\br+\delta)=& &\frac{1}{2}
    \sum_{n} J_2(\br,\br+\delta)\tanh \frac{E_n}{2k_B T}   \nonumber \\
    & &\times(u^n_{\br,\alpha}v^{n*}_{\br+\delta,\alpha}
    +v^{n*}_{\br,\alpha}u^n_{\br+\delta,\alpha}),  \label{eq:selfcon_nn}
\end{eqnarray}
and
\begin{eqnarray}
    \langle n_{\alpha,\br}\rangle &=& \langle c^\dagger_{\br\alpha}c_{\br\alpha}\rangle \nonumber \\
    &=& \sum_{n}|v^n_{\br\alpha}|^2[1-f(E_n)]+\sum_n|u^n_{\br\alpha}|^2f(E_n),  \label{eq:selfcon_ne}
\end{eqnarray}
where $f(E)$ is the Fermi distribution function. We have studied the model for a few sets of $J_1,J_2$ (values in a clean system), and a wide range of isovalent doping percentage $0<x<1$ on square lattices of sizes up to $N=32\times 32$. We always perform our computations on finite lattice sites with periodic boundary conditions. For a given (quenched) disorder configuration, we obtain the resultant quasiparticle spectrum by repeatedly diagonalizing the BdG equation~(\ref{eq:bdg}) after each iteration of the pairing amplitudes according to self-consistency conditions (\ref{eq:selfcon_nn}) and (\ref{eq:selfcon_ne}) until sufficient accuracy is achieved (e.g., the relative error of the pairing amplitudes is less than 0.01\%).

\section{Modulation of Pairing Symmetry with Bond Disorder}

\subsection{Pairing symmetry and local pairing amplitudes}
\label{sec:pairing}

\begin{figure*}[!htb]
	\begin{center}
        \includegraphics[height=2.8in]{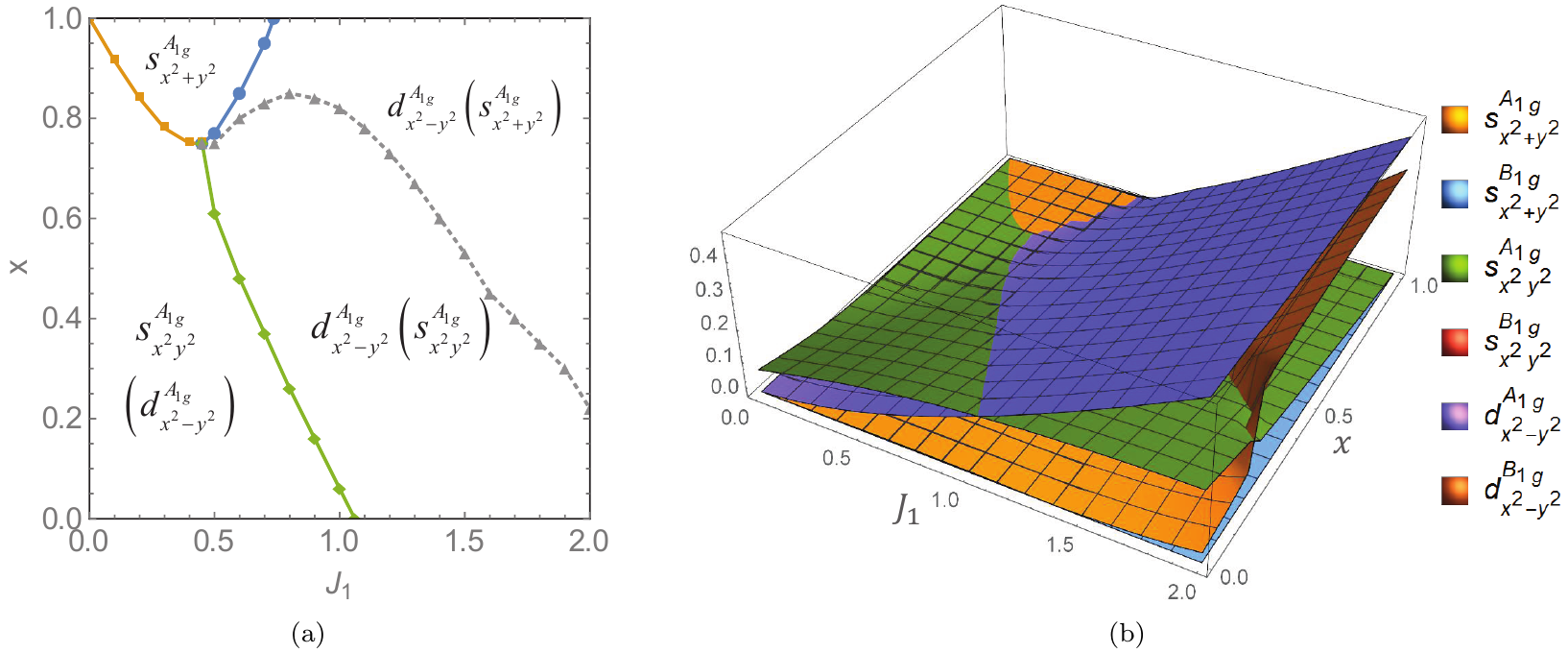}
		\caption{{(a) Schematic zero-temperature phase diagram as a function of $J_1$ and disorder $x$, given $J_2=|t_1|$ on a $32\times 32$ square lattice. The green line marks the boundary between the dominant $d_{x^2-y^2}^{A_{1g}}$-wave pairing region and $s_{x^2y^2}^{A_{1g}}$-wave pairing one, while the dashed line marks the boundary between the subdominant $s_{x^2y^2}^{A_{1g}}$-wave pairing region and $s_{x^2+y^2}^{A_{1g}}$-wave pairing one; the subdominant pairing symmetries are shown within the parentheses. (b) The position and disorder averaged pairing amplitudes for each pairing symmetry as a function of $J_1$ and $x$.}
			\label{fig:phaseJx}}
	\end{center}
\end{figure*}

\begin{figure}[ht]
    \begin{center}
        \includegraphics[width=3.3in]{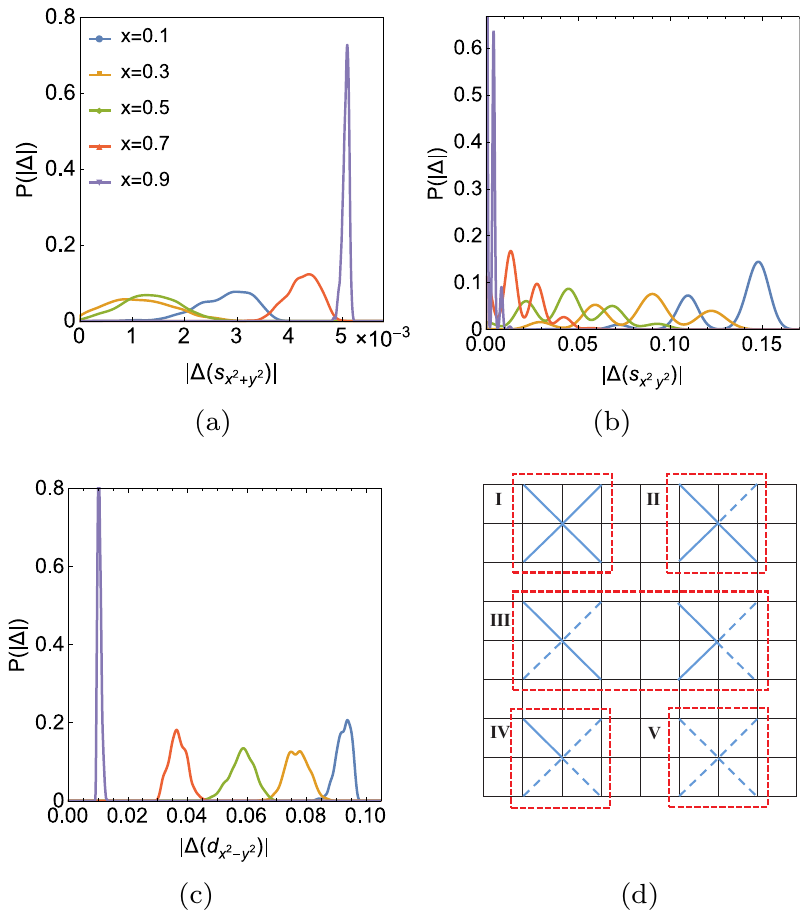}
        \caption{{Distribution of the local pairing amplitudes $P\left(|\Delta(\br)|\right)$ for various disorder $x$ at zero temperature, with $J_1=0.7$ and $J_2=1$. (a), (b), and (c) correspond to $s_{x^2+y^2}$-,
        $s_{x^2y^2}$-, and $d_{x^2-y^2}$-wave pairing amplitudes, respectively. (d) Five possible disordered configurations for a local lattice site. The dashed lines denote the NNN bonddefect ($J_2=0$).}
        \label{fig:distribution}}
    \end{center}
\end{figure}

The zero-temperature phase diagram of the $t$-$J_1$-$J_2$ model in the context of iron-based superconductors has been well studied in the clean limit, $x=0$.\cite{seo2008,QimiaoSi2010} There are four spatial symmetries for the singlet pairing by decoupling $J_1$ and $J_2$ interactions: $s_{x^2y^2}$, $s_{x^2+y^2}$, $d_{x^2-y^2}$, and $d_{xy}$. Given the two-orbital nature of our model and $D_{4h}$ point group symmetry (when taking one iron per unit cell), every spatial symmetry may even belong to different irreducible representations. For instance, one can have $s_{x^2y^2}^{A_{1g}}$ (also called $s_{\pm}$ wave) and $s_{x^2y^2}^{B_{1g}}$, or $d_{x^2-y^2}^{A_{1g}}$ and $d_{x^2-y^2}^{B_{1g}}$.

In the presence of disorder ($x\neq 0$), the pairing amplitudes could be no longer uniform and are likely to be inhomogeneous in a self-consistent manner. Therefore, we define the intraorbital spin-singlet pairing amplitude on an NN bond as (similarly for an NNN bond pairing)
\begin{eqnarray}
\Delta_{\alpha}\left(\br,\br+\delta\right) = -\frac{J_1}{2}\langle c_{\br,\alpha,\downarrow}c_{\br+\delta,\alpha,\uparrow} - c_{\br,\alpha,\uparrow}c_{\br+\delta,\alpha,\downarrow} \rangle,
\end{eqnarray}
and three dominant local pairing amplitudes with appropriate pairing symmetries all in $A_{1g}$ irreducible representation,
\begin{eqnarray}
\Delta_{s_{x^2y^2}}(\br) &=& \frac{1}{8}\sum_{\alpha\delta^\prime}\Delta_{\alpha}\left(\br,\br+\delta^\prime\right), \nonumber \\
\Delta_{s_{x^2+y^2}}(\br) &=& \frac{1}{8}\sum_{\alpha\delta}\Delta_{\alpha}\left(\br,\br+\delta\right), \nonumber \\
\Delta_{d_{x^2-y^2}}(\br) &=& \frac{1}{8}\sum_{\alpha}(-1)^{\alpha}[\Delta_{\alpha}\left(\br,\br+\hat{x}\right)-\Delta_{\alpha}\left(\br,\br+\hat{y}\right) \nonumber \\
&+& \Delta_{\alpha}\left(\br,\br-\hat{x}\right)-\Delta_{\alpha}\left(\br,\br-\hat{y}\right)],
\label{eq:localpairing}
\end{eqnarray}
where $\delta=\pm\hat{x},\pm\hat{y}$ and $\delta^\prime=\pm\hat{x}\pm\hat{y}$. When determining which pairing symmetry is most dominant for a given disorder proportion $x$, we take an average over the whole lattice positions and disorder configurations for each local pairing amplitude shown in Eq.~(\ref{eq:localpairing}).

To consider the effects of the bond disorder, we first plot the zero-temperature $x$ vs $J_1$ phase diagram in Fig.~\ref{fig:phaseJx}(a) with corresponding averaged pairing amplitudes in Fig.~\ref{fig:phaseJx}(b), where $J_2$ is fixed to be $|t_1|$. One of the most important observations is that as long as $J_1\lesssim J_2$, the pairing symmetry of the superconducting ground state would be modulated from an $s_\pm$-wave to a $d$-wave pairing symmetry when the disorder $x$ grows greater than certain critical $x_c$; on the contrary, as $J_1\gtrsim J_2$ the $d$-wave symmetry would dominate no matter what $x$ is.

Note that all the phases appear in the phase diagram belong to the $A_{1g}$ irreducible representation. Thus, in a strict sense, there should be no sharp phase transitions in our disordered system. The crossover boundary between phases, the green line, represents the degeneracy of the averaged pairing amplitudes between the {\it dominant} $s_{x^2y^2}^{A_{1g}}$ and $d_{x^2-y^2}^{A_{1g}}$ symmetries, while the dashed line represents the {\it subdominant} line between $s_{x^2y^2}^{A_{1g}}$ and $s_{x^2+y^2}^{A_{1g}}$ symmetries, as shown in Fig.~\ref{fig:phaseJx}(b). The pairing amplitudes belonging to the $B_{1g}$ irreducible representation always have a smaller
weight than those of $A_{1g}$'s. So, every pairing symmetry in this work will refer to the $A_{1g}$ irreducible representation and we will omit the superscript hereafter.

When $x=0$ (1), the pairing amplitudes for various pairing symmetry channels are {\it uniform} with $\Delta_{s_{x^2y^2}}(\br)=0.1585 > \Delta_{d_{x^2-y^2}}(\br)=0.0991$ [$\Delta_{d_{x^2-y^2}}(\br)=0.0194 >\Delta_{s_{x^2+y^2}}(\br)=0.0049$], given $J_1=0.7$ and $J_2=1$ in the system. But, as we have mentioned earlier in this subsection, the pairing amplitudes are not necessary so when $0<x<1$. In Figs.~\ref{fig:distribution}(a)-(c), we provide a statistical distribution of the local pairing amplitudes $P(|\Delta(\br)|)$ for several disorders $x$ in different pairing symmetry channels at zero temperature. When $x=0.1$ or $x=0.9$, each distribution function shows a major sharp peak, indicating the uniformity of the system. As $x$ grows from 0.1, the major peak becomes lower and broader ($|\Delta|$ are spread in a wider range of values), indicating the nonuniformity of the system which reaches its maximum at $x=0.5$. Further increasing $x$, the distribution is basically reversed and the amplitudes become more uniform again.

The behavior we found here as $x$ increases is qualitatively similar to that seen in a usual $s$-wave superconductor from weak to strong site-impurity disorder.\cite{Ghosal2001} (AG theory breaks down.) However, there exists a sharp feature not seen in the simple site-impurity case: For $s_{x^2y^2}$-wave pairing symmetry [see Fig.~\ref{fig:distribution}(b)], the distribution function shows a few extra subpeaks except for the major one for every given $x$. This is a unique property which distinguishes $s_{x^2y^2}$-wave pairing from the other two pairing channels [see Figs.~\ref{fig:distribution}(a) and \ref{fig:distribution}(c)]. In fact, this feature may be physically understood as follows. First, the $s_{x^2y^2}$-wave pairing originates from the $J_2$ exchange term, not from the $J_1$ term. Second, since the disorder is introduced by randomly taking away the NNN ($J_2$) bonds, there should be five different local disorder configurations, as illustrated in Fig.~\ref{fig:distribution}(d). Therefore, in a macroscopic system, all five configurations should in principle be reflected on the distribution function in terms of the peaks observed in Fig.~\ref{fig:distribution}(b). (Sometimes the distribution peak corresponding to either the configuration I or V is too small to be seen.)

\subsection{Density of states and energy gap}

\begin{figure}[ht]
	\begin{center}
        \includegraphics[height=2.5in]{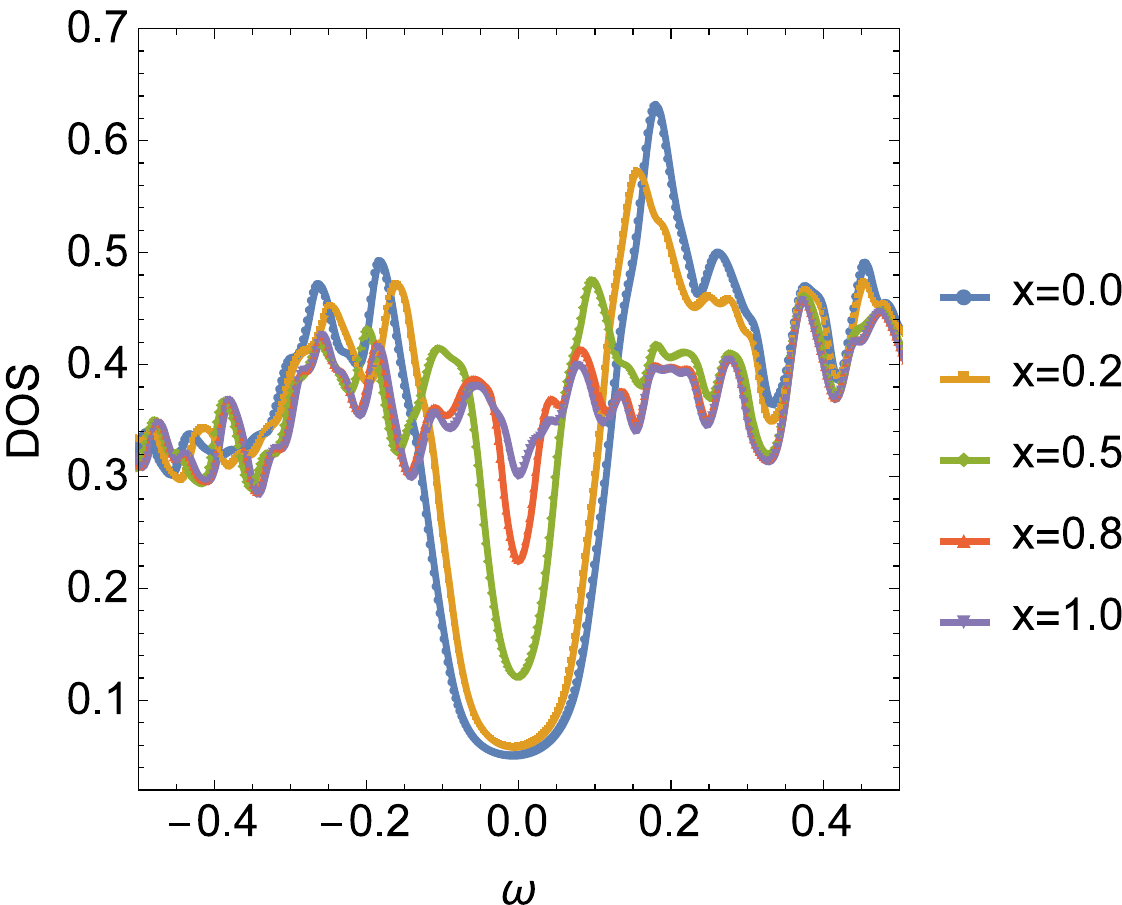}
		\caption{{Density of states for various disorder proportions $x$ at zero temperature, with $J_1=0.7$ and $J_2=1$. Each DOS for a given $x$ has been averaged over position and 30 disorder configurations.}
			\label{fig:DOS}}
	\end{center}
\end{figure}

\begin{figure}[ht]
    \begin{center}
        \includegraphics[height=2.5in]{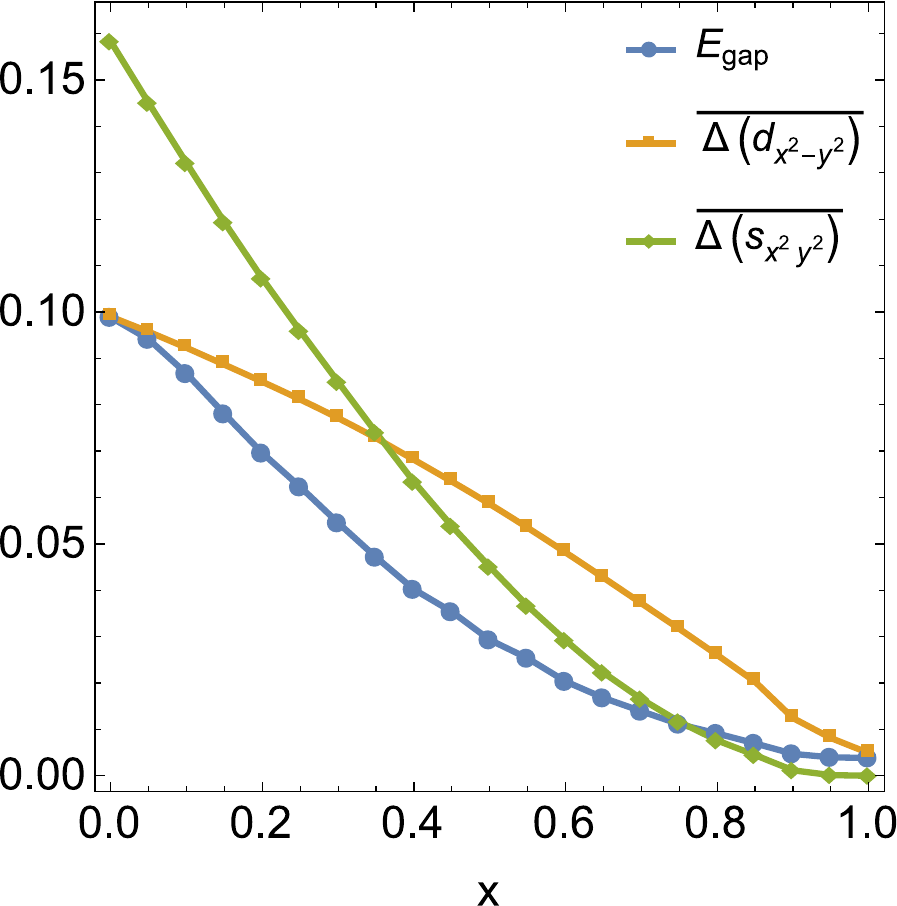}
        \caption{{Quasiparticle gap $E_{gap}$ and position/disorder averaged pairing amplitudes $\overline{\Delta}_{s_{x^2y^2}}$ and $\overline{\Delta}_{d_{x^2-y^2}}$ as a function of $x$ with $J_1=0.7$ and $J_2=1$. }
        \label{fig:gap}}
    \end{center}
\end{figure}

One simple way to demonstrate a possible modulation of pairing symmetry as $x$ increases is to calculate the electron density of states (DOS), which is expressed as
\begin{eqnarray}
    N\left(\omega\right) = \frac{1}{N}\sum_{\br\alpha n}\left[|u_{\alpha\br}^{n}|^{2}\delta\left(E_{n}-\omega\right) + |v_{\alpha\br}^{n}|^{2}\delta\left(E_{n}+\omega\right)\right], \nonumber \\
\end{eqnarray}
where the $\delta$ function has the form of Cauchy-Lorenz distribution function, $\delta(x)=\gamma/\pi(x^2+\gamma^2)$, with an infinitesimal scale parameter $\gamma=0.008$.

Fig.~\ref{fig:DOS} shows the DOS for different $x$ at zero temperature in a $56\times 56$ lattice. Each data curve is obtained by averaging over the whole lattice positions as well as 30 disorder configurations at any given $x$. With increasing disorder $x$, the superconducting gap around $\omega=0$ decreases and evolves from U-shape like to more V-shape like, indicating a modulation of the pairing symmetry from $s_{x^2y^2}$ wave to $d_{x^2-y^2}$ wave. In addition, the coherence peaks are slowly smeared out, behaving similarly to the case with impurity disorder.\cite{Ghosal2001}

The evolution of the gap in the DOS result may be complemented by the quasiparticle energy gap $E_{gap}$, which is defined as the smallest positive eigenvalue of the BdG Hamiltonian in Eq.~(\ref{eq:bdg}). As shown in Fig.~\ref{fig:gap}, $E_{gap}$ decreases when $x$ increases and it also evolves in the same trend as (position and disorder averaged) $\overline{\Delta}_{s_{x^2y^2}}$; otherwise, if there were no $\Delta_{s_{x^2y^2}}$ component in the SC state, $E_{gap}$ should be ideally zero due to nodal quasiparticle excitations from $d$-wave pairing. Note that when $x\gtrsim 0.75$, $E_{gap}$ is slightly greater than $\overline{\Delta}_{s_{x^2y^2}}$, which is due to the presence of $\overline{\Delta}_{s_{x^2+y^2}}$ in the system (not shown in Fig.~\ref{fig:gap}). Therefore, $E_{gap}$ appears to be dictated mainly by $\overline{\Delta}_{s_{x^2y^2}}$ in a certain way. The subtle feature of it could be further revealed by the physics we will discuss next.

\subsection{Formation of superconducting islands}

\begin{figure}[ht]
    \begin{center}
        \includegraphics[width=3.3in]{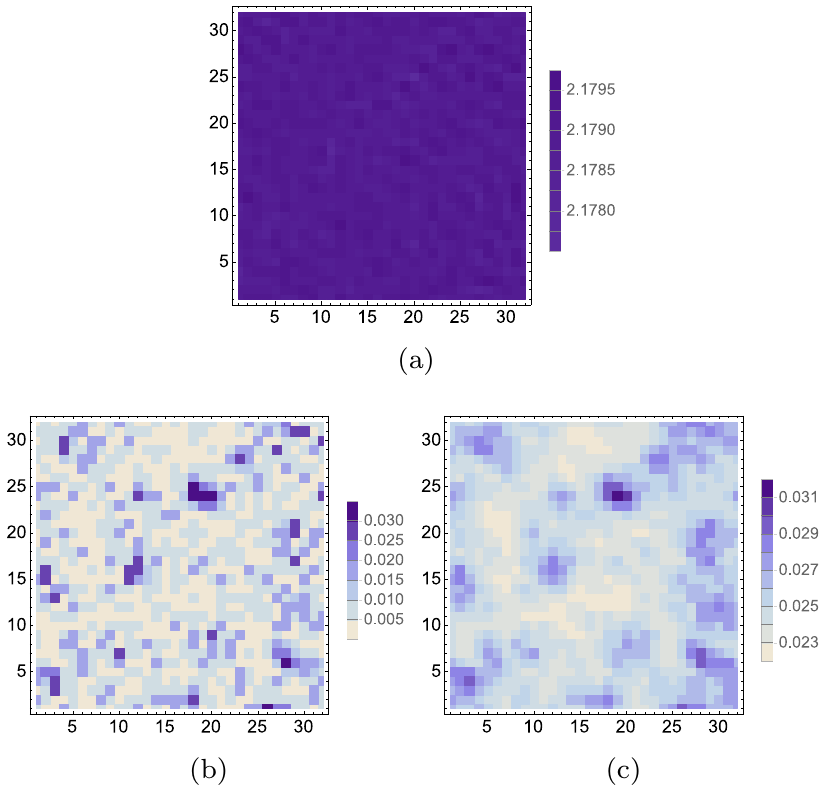}
        \caption{{(a) Electron density on a $32\times 32$ lattice for a given disorder realization with $x=0.8$ at zero temperature. (b) and (c) correspond to two dominant $s_{x^2y^2}$- and $d_{x^2-y^2}$-wave pairing
        channels. The darker the larger density/amplitude would be. }
        \label{fig:localization}}
    \end{center}
\end{figure}

\begin{figure}[ht]
    \begin{center}
        \includegraphics[width=3.3in]{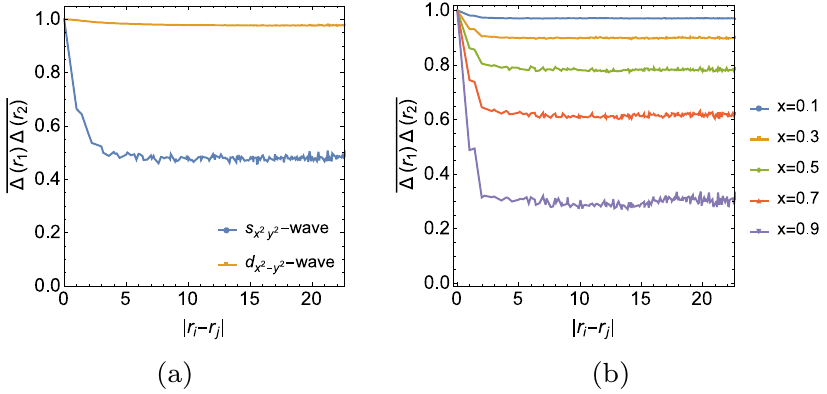}
        \caption{{(a) The disorder-averaged correlation functions $\overline{\Delta(\br_1)\Delta(\br_2)}$ for $s_{x^2y^2}$- and $d_{x^2-y^2}$-wave pairing amplitudes as a function of the distance $r_{ij}=|\br_1-\br_2|$ with
        $x=0.8$. (b) $\overline{\Delta(\br_1)\Delta(\br_2)}$ of $s_{x^2y^2}$-wave pairing for various $x$. Both (a) and (b) are averaged over 15 random disorder configurations and both are scaled by
        $\overline{\Delta^2(\br_1)}$.}
        \label{fig:correlation}}
    \end{center}
\end{figure}

In our study, local pairing amplitudes are self-consistently calculated at every lattice site by using BdG theory for a given disorder realization. In particular, as mentioned in Sec.~\ref{sec:pairing}, the distribution of $s_{x^2y^2}$-wave pairing amplitudes seems to be correlated with local disorder configurations. Usually this should be contrasted with the electron density distribution. Therefore, in Figs.~\ref{fig:localization}(a), \ref{fig:localization}(b), and \ref{fig:localization}(c), we plot the electron density $n(\br)=\sum_{\alpha\sigma}\langle n_{\alpha,\br,\sigma}\rangle$ and spatial variation of $s_{x^2y^2}$-wave and $d_{x^2-y^2}$-wave pairing amplitudes, respectively, for a given disorder realization with $x=0.8$ at zero temperature. Obviously, the electron density $n(\br)$ is nearly homogeneous and uncorrelated with the given bond disorder. However, the $s_{x^2y^2}$-wave pairing amplitudes are inhomogeneous and tend to form some $s$-wave SC ``islands'' on top of a relatively less inhomogeneous $d$-wave SC ``sea,'' even though there is no intrinsic correlation between disordered bonds. Note that there is also a small clustering tendency for $d_{x^2-y^2}$-wave pairing amplitudes, which is likely to be induced by an $s$-wave piece.

To further confirm the formation of $s$-wave SC ``islands'' without a strong correlation with a certain disorder configuration, we display the disorder-averaged correlation functions $\overline{\Delta(\br_1)\Delta(\br_2)}$ for $s_{x^2y^2}$- and $d_{x^2-y^2}$-wave pairing amplitudes as a function of the distance $r_{ij}=|\br_1-\br_2|$ at $x=0.8$ in Fig.~\ref{fig:correlation}(a). The $s$-wave component shows a clear structure on a scale of two to three lattice constants, while the $d$-wave counterpart shows almost no structure. Note that the coherence length in a clean $s_{x^2y^2}$-wave SC in this model can be estimated by $v_F/\Delta\sim 4$. Moreover, the clustering tendency for the $s$-wave component persists as $x$ increases as can be seen in Fig.~\ref{fig:correlation}(b).

Additionally, it is worth noticing that for the first few lowest quasiparticle excited states, each corresponding wave function distribution, $|\Psi(\br)|^2$, appears to be a propagating mode with a definite momentum along the diagonal direction; along each wave front (e.g., of the largest amplitude), the probability distributes almost uniformly but suppresses around $s$-wave SC ``islands.'' Such variation in $|\Psi(\br)|^2$ suggests its nature as a gapped mode. Thus, this feature might well explain why $E_{gap}>0$, as found in Fig.~\ref{fig:gap}, is dictated by the $s$-wave pairing channel in our inhomogeneous SC system.

\subsection{Superfluid density and critical temperature}
\label{sec:Tc}

The temperature dependence of the superfluid density $\rho_s(T)$ is usually viewed as a good indicator to determine whether a superconductor is ``unconventional'' or not. For instance, $\rho_s(T)$, which is measurable through the magnetic penetration depth $\lambda^{-2}(T)$, exhibits an exponential behavior in $T$ in a clean (hole-doped) 122 compound, Ba$_{1-x}$K$_x$Fe$_2$As$_2$, suggesting a fully gapped SC;\cite{Hashimoto2009} while $\rho_s(T)$ behaves like a power law in an electron-doped Ba(Fe$_{1-x}$Co$_x$)$_2$As$_2$, indicating the presence of nodes or something unconventional.\cite{Gordon2009,Gordon2010,H.Kim2010}

Moreover, the qualitative behavior of $\rho_s(T)$ gives a useful hint to understand (1) how disorder of certain type may affect SC (phase rigidity) at low temperatures\cite{Franz1997,Ghosal2001} and (2) what type of fluctuation is essential near critical temperature $T_c$.\cite{Zuev05} Although the answer to question (2) is beyond our mean-field study, we can still use it to obtain $T_c$ in a self-consistent manner when inhomogeneous pairing is inevitable and not negligible.

We generalize the linear-response approach in Refs.~\onlinecite{Scalapino1992} and \onlinecite{Scalapino1993} for a multiorbital case to obtain the superfluid density. Considering a weak vector potential $A_x(\br,t)$ along the $x$ direction, the hopping terms in Eq. (\ref{eq:h0}) are modified by the Peierls phase factors $-e^{ieA_x(\br)}$. Expanding the phase factors up to second order, we obtain
\begin{eqnarray}
    H_t^A = H_t - \sum_{\br}\left[ej_x^p(\br)A_x(\br)+\frac{1}{2}e^2k_x(\br)A_x^2(\br)\right],
    \label{eq:hA}
\end{eqnarray}
with the paramagnetic current
\begin{eqnarray*}
    j_x^p(\br) = i\sum_{s\sigma}\sum_{\alpha\beta} t_{\br,\br+s}^{\alpha\beta} \left(c^{\dagger}_{\br+s,\beta\sigma}c_{\br\alpha\sigma} -  c^{\dagger}_{\br\alpha\sigma}c_{\br+s,\beta\sigma}\right),
    \label{eq:jp}
\end{eqnarray*}
and the kinetic energy associated with $x$ direction
\begin{eqnarray*}
    k_x(\br) = -\sum_{s\sigma}\sum_{\alpha\beta} t_{\br,\br+s}^{\alpha\beta} \left(c^{\dagger}_{\br+s,\beta\sigma}c_{\br\alpha\sigma} +  c^{\dagger}_{\br\alpha\sigma}c_{\br+s,\beta\sigma}\right).
    \label{eq:kx}
\end{eqnarray*}
According to the linear response theory,
\begin{eqnarray}
    \rho_s = \langle -k_x \rangle - \Lambda_{xx}\left(q_x=0,q_y \rightarrow 0,i\omega=0\right),
    \label{eq:rho}
\end{eqnarray}
where $\langle -k_x \rangle$ is the kinetic energy along $x$ direction, and
\begin{eqnarray}
    \Lambda_{xx}\left(\mathbf{q},i\omega_m\right) = \frac{1}{N^2}\int_0^{\beta}d\tau e^{i\omega_m\tau}\langle j_x^p\left(\mathbf{q},\tau\right) j_x^p\left(-\mathbf{q},0\right) \rangle.   \nonumber \\
    \label{eq:rhoq}
\end{eqnarray}
Note that in our disordered model, we calculate the current-current correlation function in real space,
\begin{eqnarray}
    \Lambda_{xx}\left(\br_1,\br_2,i\omega_m\right) = \frac{1}{N}\int_0^{\beta}d\tau e^{i\omega_m\tau}\langle j_x^p\left(\br_1,\tau\right) j_x^p\left(\br_2,0\right) \rangle.   \nonumber \\
    \label{eq:rhor}
\end{eqnarray}
After disorder averaging, the translational invariance may be recovered and hence we set $\Lambda_{xx}\left(\br_1,\br_2,i\omega\right) = \Lambda_{xx}\left(\br_1-\br_2,i\omega\right)$, followed by a Fourier transform to $\bq$ space,
\begin{eqnarray}
    \Lambda_{xx}\left(\mathbf{q},i\omega\right) = \frac{1}{N^2}\sum_{\br_1\br_2}e^{-i\bq\cdot(\br_1-\br_2)} \Lambda_{xx}\left(\br_1,\br_2,i\omega\right).
    \label{eq:rhok}
\end{eqnarray}

\begin{figure}[ht]
    \begin{center}
        \includegraphics[width=3.2in]{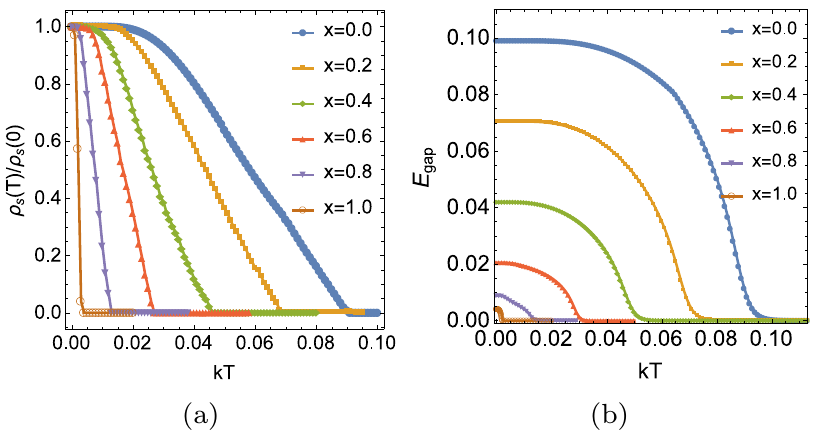}
        \caption{{(a) Temperature dependence of superfluid density $\rho_s(T)$ and energy gap $E_{gap}$ for various disorder proportions $x$. }
        \label{fig:sfgap}}
    \end{center}
\end{figure}

\begin{figure}[ht]
    \begin{center}
        \includegraphics[height=2.5in]{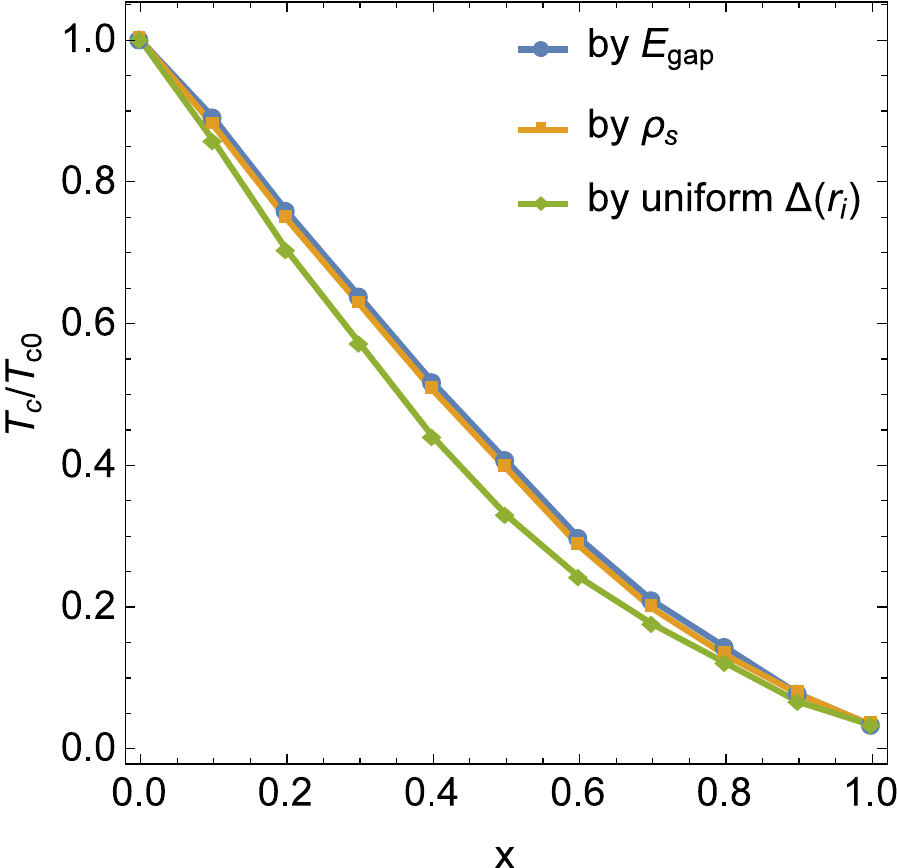}
        \caption{Critical temperature $T_c$ as a function of $x$, obtained by quasiparticle gap minimum $E_{gap}$, superfluid density $\rho_s$, and the AG-type calculation with enforced uniform $\Delta(\br_i)$,
        respectively.
        \label{fig:Tc}}
    \end{center}
\end{figure}

In Fig.~\ref{fig:sfgap}(a), we show the temperature dependence of $\rho_s(T)$, normalized by $\rho_s(0)$, at various $x$. At any given $x$, $\rho_s(T)$ deviates from $\rho_s(0)$ exponentially at low temperatures and then decreases until reaching zero at a critical $T$ (called $T_c$). This can be contrasted by the minimum quasiparticle energy $E_{gap}$ as a function of $T$ depicted in Fig.~\ref{fig:sfgap}(b) with a similar trend. Such an exponential behavior strongly suggests the SC ground state of the system is fully gapped.

A few remarks are worth mentioning here. First, unlike in the case of site-impurity disorder, $\rho_s(0)$ is basically not suppressed as the disorder strength $x$ increases, indicating that the phase coherence for electrons is almost unaffected.
Second, at $x>x_c$ even though the $d$-wave component is dominant, there could be no gapless Bogoliubov quasiparticles, as indicated in Fig.~\ref{fig:sfgap}(b) at $T<T_c$.
Third, the $T_c$'s determined by $\rho_s(T)$ and $E_{gap}(T)$ are not necessarily the same. In fact, $E_{gap}$ usually overestimates $T_c$ slightly in our study (see Fig.~\ref{fig:Tc}). This is because a system may have certain local pairing amplitudes while completely losing the phase coherence among them. Therefore, $T_c$ determined by $\rho_s$ would be more reliable.

Typically the standard AG theory provides an estimation of the suppressed $T_c$ in a disordered superconductor. The suppressed $T_c$, $\Delta T$, is basically proportional to the disorder scattering rate, $x/N(0)$, times a constant of $\mathcal{O}(1)$. The essential step in its derivation is to replace $\Delta(\br)$ by the spatially averaged one. Therefore, we also calculate the quasiparticle gap with an enforced uniform order parameter $\Delta(\br_i)$ in the self-consistent mean-field equations and summarize all of our obtained $T_c$ in Fig.~\ref{fig:Tc}. It is important to observe that $E_{gap}$-determined $T_c$ via the inhomogeneous solution in the self-consistent manner agrees well with the $\rho_s$-determined one, but both are away from the AG-type $T_c$ when $0<x<1$. This strongly implies the breakdown of the conventional AG-type theory for the bond-disordered
system.

\section{System without $J_1$}

\begin{figure}[ht]
    \begin{center}
        \includegraphics[width=3.3in]{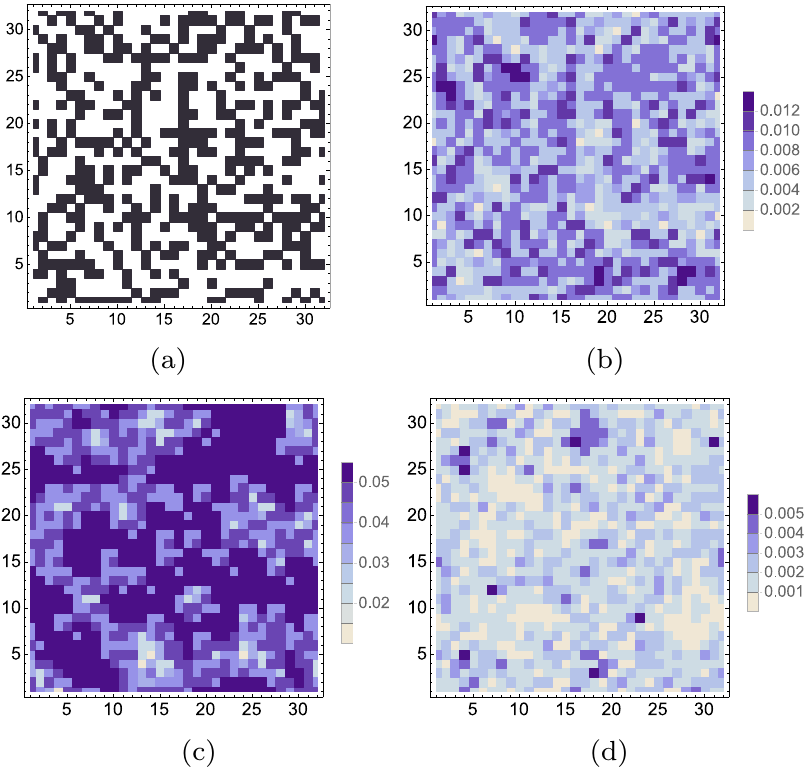}
        \caption{(a) A disordered pattern with $x=0.4$. The black regions mark the points with $J_2=0$. (b) The spatial distribution of local pairing amplitudes corresponding to (a). The darker purple color indicates the regions with larger pairing amplitudes. (c) and (d) are another two spatial distributions of local pairing amplitudes with $x=0.1$ and $x=0.7$, respectively. The darker purple color has the same meaning explained in (b)}.
        \label{fig:localizationJ1}
    \end{center}
\end{figure}

\begin{figure}[ht]
    \begin{center}
        \includegraphics[height=2.5in]{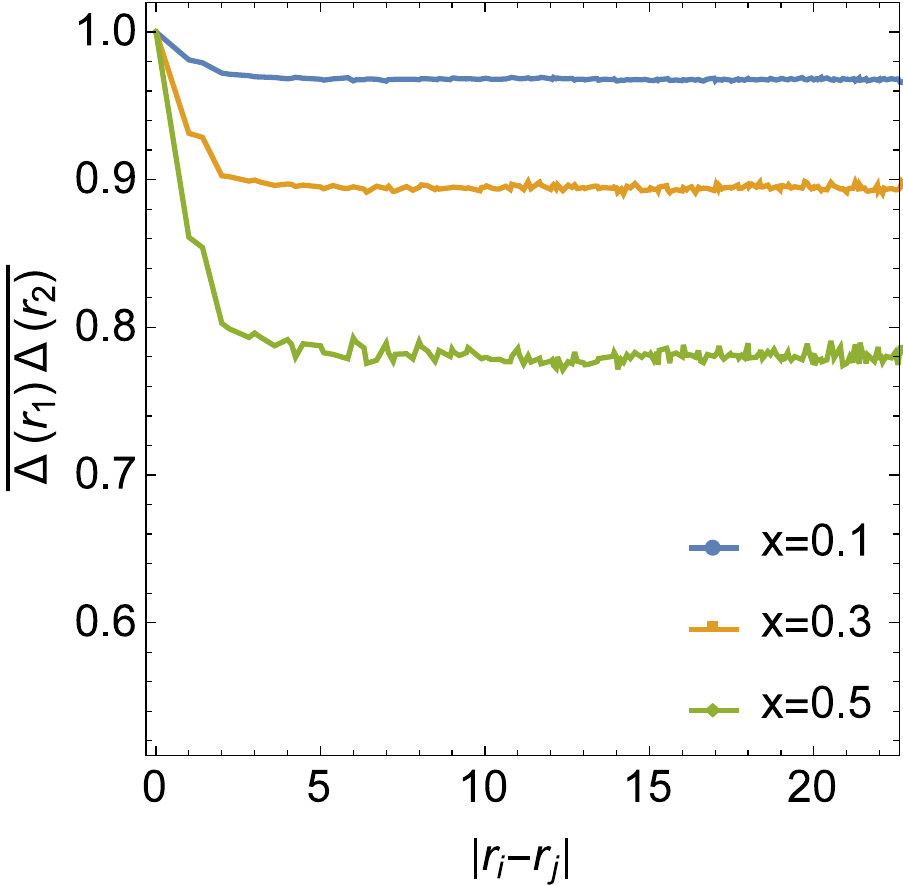}
        \caption{The correlation function $\overline{\Delta(\br_1)\Delta(\br_2)}$ for $s_{x^2y^2}$-pairing amplitudes as a function of the distance $r_{ij}=|\br_1-\br_2|$ with different $x$, averaged by 15 configurations
        and scaled by $\overline{\Delta^2(\br_1)}$.}
        \label{corr_sx2y2J1}
    \end{center}
\end{figure}

\begin{figure}[ht]
	\begin{center}
		\includegraphics[height=2.5in]{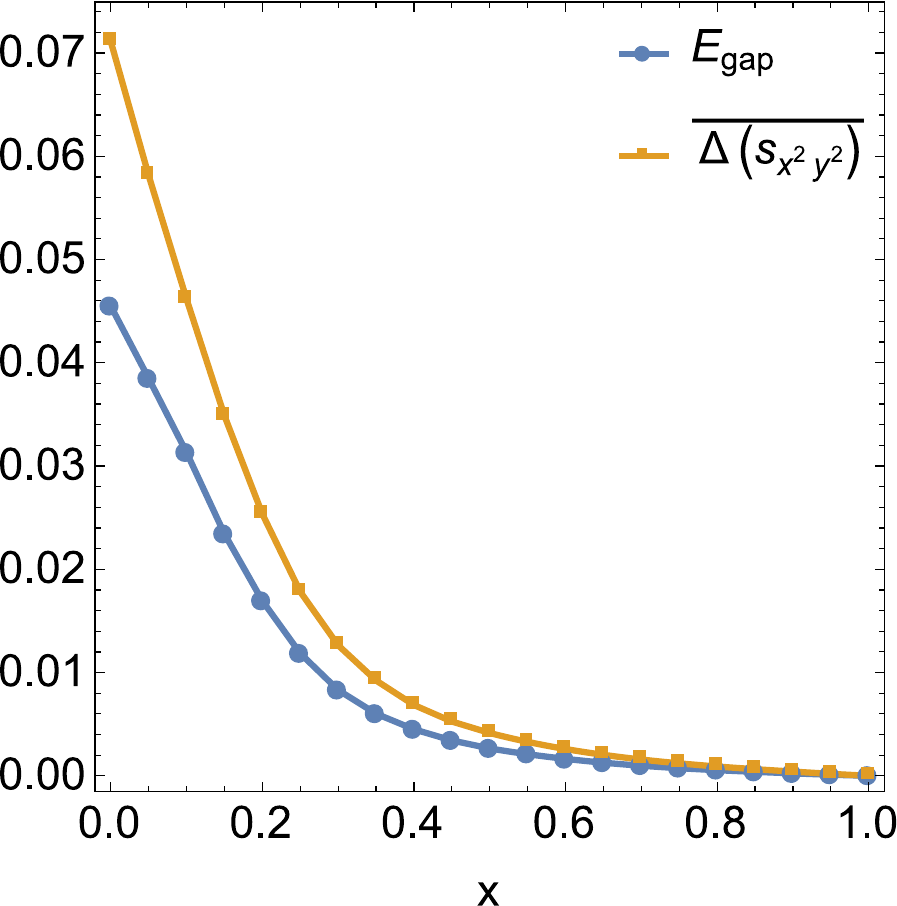}
        \caption{Quasiparticle gap $E_{gap}$ and position/disorder averaged pairing amplitudes $\overline{\Delta}_{s_{x^2y^2}}$  as a function of $x$ with $J_1=0$ and $J_2=1$.
		  \label{fig:gapJ1}}
	\end{center}
\end{figure}

\begin{figure}
	\begin{center}
		\includegraphics[height=2.5in]{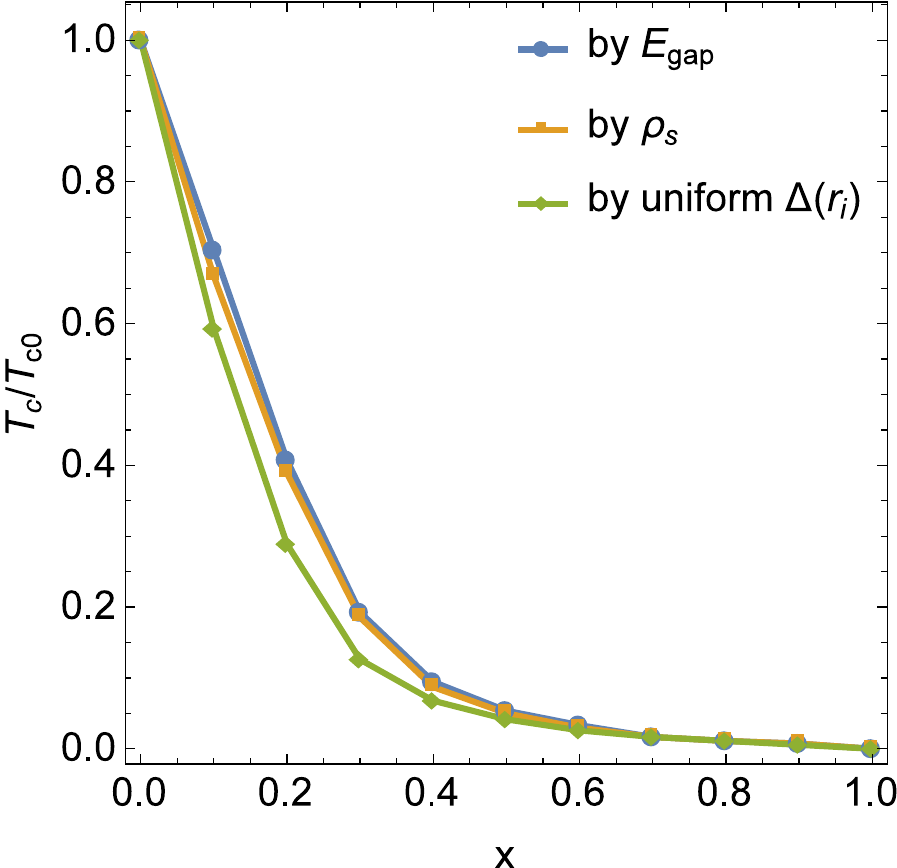}
        \caption{Superconducting critical temperature $T_c$ as a function of $x$, with $J_1=0, J_2=1$.
        \label{fig:TcJ1}}
	\end{center}
\end{figure}

We have studied a $J_2$-bond disordered $t$-$J_1$-$J_2$ model in the previous sections. However, the presence of $J_1$ terms may couple with $J_2$ terms in a complicated way and features purely from the bond disorder could be still obscure. Since this type of disorder is interesting in its own right, in this section, we repeat previous calculations by setting $J_1=0$.

Focusing on $s_{x^2y^2}$ wave only, Fig.~\ref{fig:localizationJ1}(b) shows the distribution of the spatial pairing amplitudes at zero temperature with its corresponding disorder realization at $x=0.4$ depicted in Fig.~\ref{fig:localizationJ1}(a). The trend to form ``superconducting islands'' is clear [see Figs.~\ref{fig:localizationJ1}(c)$\rightarrow$(b)$\rightarrow$(d)]: As $x\gtrsim 0.5$, ``superconducting islands'' are formed while gradually becoming smaller until $x$ approaches 1. These ``islands'' are instead separated from one another by a metallic ``sea'' (with negligible $\Delta(\br)$). This can be more accurately confirmed by the disorder-averaged correlation function $\overline{\Delta(\br_1)\Delta(\br_2)}$ for various $x$, as shown in Fig.~\ref{corr_sx2y2J1}.

We next examine the quasiparticle minimum gap $E_{gap}$ as a function of the disorder proportion $x$ at zero temperature, as illustrated in Fig.~\ref{fig:gapJ1}. Clearly, $E_{gap}$ follows the same trend as $\overline{\Delta}_{s_{x^2y^2}}$. Note that even though the system tends to form the ``superconducting islands'' as $x$ grows, the superfluid stiffness $\rho_s$ does not vanish until $x=1$ (see Fig.~\ref{fig:TcJ1}). This is because, within our BdG formalism, we ignored the possible phase fluctuations between SC islands, which can frustrate Cooper pairs to form a coherent superconducting state. Thus, in principle, one may not rule out the occurrence of a superconductor-metal transition in this purely academic model. Moreover, in Fig.~\ref{fig:TcJ1} the $T_c$'s determined by $E_{gap}$ and by the superfluid density $\rho_s$ are consistent with each other, while the AG-type calculation is again shown to underestimate $T_c$, reflecting the importance of the inhomogeneous pairing amplitudes.

\section{Discussion and conclusion}
Before concluding our work, there are a few additional remarks worth mentioning here, which are relevant to isovalent doping in iron-based SCs.
First, as we have mentioned in Sec. II, in the strong-coupling picture the leading effect of the substitution of P for As reduces seriously the exchange interaction $J_2$ (presumably superexchange type).\cite{j2suppression} However, a number of subleading effects are also anticipated. For instance, one might consider including the weakened NN exchange interaction $J_1$ and the presence of short-range impurity potentials. Both of them are basically pair breakers: The former suppresses the $d$-wave superconductivity, while the latter could erase the phase information or even insulate originally superconducting regions. As long as the ratio of the impurity potential strength to $J_1$ is much less than 1, our results shown in the previous sections may remain valid qualitatively.

Second, although our prediction on the modulation of the pairing symmetry may be applicable for the isovalent-doping 1111 iron-based SCs at finite temperatures, where LaOFeAs$_{1-x}$P$_x$ ($x=1$) has been reported to have gap nodes,\cite{Fletcher2009,Hicks09} our theory cannot sufficiently describe 122 systems. This is because we have so far ignored the magnetic and orbital fluctuations, which are believed to play important roles in exhibiting either an antiferromagnetic order or a nematic state.\cite{Liang15} We will refer this more generic consideration to a future study.

In summary, we studied the bond disorder effects, from ``weak'' to ``strong,'' in an unconventional superconductor described by the two-orbital $t$-$J_1$-$J_2$ model. We used the self-consistent Bogoliubov--de Gennes formulation to emphasize the necessity of the spatial inhomogeneity for the pairing amplitudes. In particular, we found that as long as $J_1\lesssim J_2$, the pairing symmetry of our model at $T=0$ can be modulated from $s_{x^2y^2}$-wave to $d_{x^2-y^2}$-wave symmetry when the ``disorder strength'' $x$ goes beyond $x_c$. This result was best presented by the electron density of states as a function of $x$ and could be partially justified by any negative evidence in experiments about probing fully gapped $s_\pm$-wave order.\cite{Tsai09,Hanaguri10,Chen11} Moreover, when $x$ is large, we observed the formation of $s_{x^2y^2}$-wave ``islands'' with length scale $\mathcal{O}(\xi)$ embedded in a $d_{x^2-y^2}$-wave ``sea,'' due to the combined pairing interaction and the $J_2$-bond disorder. As a consequence, $T_c$ determined by the superfluid density $\rho_s(T)$ is found to deviate from the predicted value by the AG theory, suggesting its insufficiency to describe the bond disorder effects.

\begin{acknowledgments}
We would like to acknowledge Fan Yang, Jiangping Hu, E. W. Carlson, and Elbio Dagotto for stimulating discussions. W.F.T. is supported in part by MOST in Taiwan under Grant No.103-2112-M-110-008-MY3 and the Thousand-Young-Talent Program of China. Y.T.K. and D.X.Y. acknowledge support from NBRPC-2012CB821400, NSFC-11574404, NSFC-11275279, NSFG-2015A030313176, Special Program for Applied Research on Super Computation of the NSFC-Guangdong Joint Fund (the second phase), Shanhai Forum, and Fundamental Research Funds for the Central Universities of China.
\end{acknowledgments}



\begin{thebibliography}{99}
\bibitem{Josephson}
In two dimensons, the interplay between localization and superconductivity due to certain types of disorder (including random many-body interactions), is usually represented by the disordered quantum XY model or
Josephson-junction array model after integrating out all fermionic degrees of freedom.
\bibitem{j2suppression}
In reality, $J_2$ would be suppressed by P substitution, but not be destroyed (diluted) completely. Thus, we have also simulated the cases with partially suppressed $J_2$: For instance, the strength of $J_2$ becomes a half
or a quarter. However, all results are qualitatively the same, indicating the validity of our simplification by using bond dilution.
\end{thebibliography}
\end{document}